\documentclass[12pt]{article}
\usepackage{mathrsfs}
\usepackage[square, numbers, compress]{natbib}
\usepackage{caption}
\usepackage{indentfirst}
\usepackage{xcolor}
\usepackage{algorithm}
\usepackage{algorithmic}
\usepackage{amsmath, amssymb}
\usepackage{amsthm}
\usepackage{multirow}
\usepackage{graphicx}
\usepackage{subcaption}
\usepackage{booktabs}
\usepackage{comment}
\usepackage[paperwidth=8.5in,paperheight=11in,
            left=1in,right=1in,top=1.0in,bottom=1.0in]{geometry}
\usepackage{titlesec}
\usepackage{rotating}
\usepackage{authblk}
\usepackage{url}
\usepackage{xr}

\title{\textbf{Bayesian Functional Analysis for Untargeted Metabolomics Data with Matching Uncertainty and Small Sample Sizes}\thanks{This manuscript has been accepted by \textit{Briefings in Bioinformatics}.}}
\author[1]{Guoxuan Ma}
\author[1,*]{Jian Kang}
\author[2,3,4,*]{Tianwei Yu}
\affil[1]{Department of Biostatistics, University of Michigan, Ann Arbor, MI 48109, USA}
\affil[2]{School of Data Science, The Chinese University of Hong Kong -- Shenzhen, Shenzhen, Guangdong 518172, China}
\affil[3]{Shenzhen Research Institute of Big Data, Shenzhen, Guangdong 518172, China}
\affil[4]{Guangdong Provincial Key Laboratory of Big Data Computing, Shenzhen 518172, China}
\affil[*]{To whom correspondence should be addressed: jiankang@umich.edu, yutianwei@cuhk.edu.cn}
\date{\vspace{-6ex}}

\def\spacingset#1{\renewcommand{\baselinestretch}%
{#1}\small\normalsize} \spacingset{1}

\def\mN{\mathrm{N}}

\begin{document}
\spacingset{1.2}
\maketitle

\begin{abstract}
Untargeted metabolomics based on liquid chromatography-mass spectrometry technology is quickly gaining widespread application given its ability to depict the global metabolic pattern in biological samples. However, the data is noisy and plagued by the lack of clear identity of data features measured from samples. Multiple potential matchings exist between data features and known metabolites, while the truth can only be one-to-one matches. Some existing methods attempt to reduce the matching uncertainty, but are far from being able to remove the uncertainty for most features. The existence of the uncertainty causes major difficulty in downstream functional analysis. To address these issues, we develop a novel approach for Bayesian Analysis of Untargeted Metabolomics data (BAUM) to integrate previously separate tasks into a single framework, including matching uncertainty inference, metabolite selection, and functional analysis. By incorporating the knowledge graph between variables and using relatively simple assumptions, BAUM can analyze datasets with small sample sizes. By allowing different confidence levels of feature-metabolite matching, the method is applicable to datasets in which feature identities are partially known. Simulation studies demonstrate that, compared with other existing methods, BAUM achieves better accuracy in selecting important metabolites that tend to be functionally consistent and assigning confidence scores to feature-metabolite matches. We analyze a COVID-19 metabolomics dataset and a mouse brain metabolomics dataset using BAUM. Even with a very small sample size of 16 mice per group, BAUM is robust and stable. It finds pathways that conform to existing knowledge, as well as novel pathways that are biologically plausible. 
\end{abstract}

\section{Introduction}
Untargeted metabolomics by liquid chromatography-mass spectrometry (LC-MS) measures small molecules in a system in an unbiased manner. It has gained increasing prominence in biomedical research for the purposes of understanding nutrition, metabolic diseases, environmental health, and cancer \cite{RN9, RN62}. 

In LC-MS metabolomics analysis, a key obstacle is the uncertainty in the matching between measured features and known metabolites \cite{RN59}. Unlike gene expression data measured by deep sequencing, LC-MS features lack direct chemical identity information. Potential matches between features and metabolites are largely based on the features' mass-to-charge ratio (m/z) and retention time (RT), often resulting in multiple potentially matched metabolites for a single feature and vice versa \cite{RN32, RN28}. Although some methods have utilized information, such as tandem mass spectrometry ($\text{MS}^2$) or feature-feature relations, to improve features-to-metabolite annotations \cite{RN32, RN4}, the uncertainty remains for most features \cite{RN33}, which posts challenges for downstream analyses and data interpretation.

The downstream analysis workflow typically involves two key tasks: the selection of metabolic features exhibiting differential abundance between sample groups and the assessment of pathway significance by considering the significance levels of features associated with each pathway \cite{RN17, RN16, RN61, RN15, RN1}. An alternative approach is to analyze the data jointly with the entire metabolic network, which avoids artificially dividing the metabolic network into pathways and utilizes the detailed connection structure between metabolites \cite{RN2, RN65}. Both suffer from the uncertainty in feature-metabolite matching.

In essence, these downstream tasks can be treated as on-network feature selection problems. A number of previous studies rely on parametric regression models for feature selection on biological networks \cite{wei2007markov, pan2010incorporating, jacob2012more, sun2014network, dona2017powerful, ren2019robust}. However, linear regression models may not capture complex associations between features and clinical outcomes effectively, potentially introducing undesirable bias \cite{zhao2014bayesian}. On the other hand, complex non-linear parametric model are more expressive but can be computationally demanding and prone to overfitting in scenarios with a large number of features and a limited number of samples, which is common in biological network analysis. Furthermore, regression models assume the presence of an outcome variable, which may not always be the case or may not align with the research focus, such as when studying biomarker expression behavior, for example, gene periodicity \cite{zhao2014bayesian, jin2022feature}. A different method is to generate a test statistic for each feature and perform feature selection using a Bayesian nonparametric approach that takes into account both the network dependencies and the test statistic \cite{zhao2014bayesian, lan2016bayesian}, which does not require parametric models nor outcome variables. \citet{jin2022feature} have extended this framework to accommodate asymmetrical null and alternative distributions and handle missing values systematically. However, these summary-statistics-based approaches, along with other regression-based methods, are designed for single layer networks (e.g., gene networks) and cannot simultaneously address both aspects in the down-stream analyses. The statistical down-stream analyses of the LC-MS metabolomics data require inferences on two-layer networks comprising observed metabolite features and the underlying metabolite network, where the matching uncertainty in between needs to be addressed. Some methods account for matching uncertainty in pathway testing by down-weighting features matched to multiple metabolites \cite{RN100tian}. However, such methods down-weight multiple-matched features evenly across all impacted pathways without inferences on which potential matching is more likely to be true. Furthermore, these methods rely on predefined pathways, ignoring the detailed metabolic network structure, and can be sensitive to the feature-level $p$-value threshold.

In this study, we propose an innovative method, Bayesian Analysis for Untargeted Metabo-lomics data (BAUM), which jointly models feature-metabolite matching and metabolic network behavior under a Bayesian semiparametric framework. We assume a summary statistic for each feature is precalculated, either in a supervised manner when clinical outcomes are available or unsupervised when they are not. This approach offers computational efficiency, avoids making parametric model assumptions, provides flexibility for both linear and non-linear relationships, and allows for analysis even when outcome variables are absent. Additionally, it can effectively manage data with small sample sizes, as it only necessitates summary statistics from observed features. Although similar concepts have found success in gene networks \cite{zhao2014bayesian, lan2016bayesian, jin2022feature}, they have not been applied to the matching uncertainty problem inherent in two-layer feature-metabolite networks. The summary statistics can be obtained by transforming $p$-values obtained by statistical tests into normally distributed statistics, where the transformation is monotone so that it is guaranteed significant features have a larger summary statistics value. We establish the connection between feature-level summary statistics and on-network latent metabolite scores while accounting for matching uncertainty through a Bayesian factor analysis model with one-hot constraints (Section \ref{sec:constrained_FA}), where the feature summary statistic is a noisy copy of the latent score for the matched metabolite. We then model the latent metabolite scores by a mixture distribution of two components, one representing clinically relevant metabolites (alternative component) and the other representing clinically irrelevant metabolites (null component), by which we control local false discovery rate (FDR) (Section \ref{sec:latent_mixture}). We assume the null component follows a centered Gaussian distribution while the alternative component follows the Dirichlet process mixture (DPM), which has been extensively discussed in Bayesian statistics \cite{antoniak1974mixtures, escobar1994estimating, neal2000markov, dunson2010nonparametric} and used in local FDR control \cite{zhao2014bayesian, lan2016bayesian, wang2019bayesian, jin2022feature}. The DPM is proved to achieve good performance on density estimation in light of its nonparametric nature, and efficient computational techniques are available, such as Gibbs sampling for stick-breaking priors \cite{ishwaran2001gibbs}. To incorporate the metabolic network information into metabolite inferences, we employ the weighted Potts prior \cite{jin2022feature} that extends the Ising prior \cite{li2008network} to assign class labels to all metabolites based on their network dependencies (Section \ref{sec:weigted_potts}). The Ising prior tends to assign similar labels to closely connected nodes on the network, making it suitable for sub-network significance analysis. For posterior computation, we develop an efficient Gibbs sampler (Section \ref{sec:posterior_inferences}), where we leverage an equivalent model representation of the DPM using the stick-breaking priors, resort to the Swendsen-Wang algorithm \cite{swendsen1987nonuniversal} for eﬃciently updating metabolite class labels, and exploit conjugacy for posterior sampling.

In several regards, BAUM is the first of its kind. Firstly, it can quantitatively evaluate which metabolite is more likely to be the true match of each feature. Secondly, BAUM is the first method to perform statistical inference directly at the metabolite level while accounting for the matching uncertainty. It can identify sub-networks within the entire metabolic network based on feature-level summary statistics, enhancing biological interpretation. Lastly, BAUM demonstrates robustness even when dealing with small sample sizes, as demonstrated in our real data analysis.

\section{Methods}\label{sec:methods}

\subsection{Overview}
\begin{sidewaysfigure}
\centering
\small
\spacingset{1}
\includegraphics[scale = 0.26]{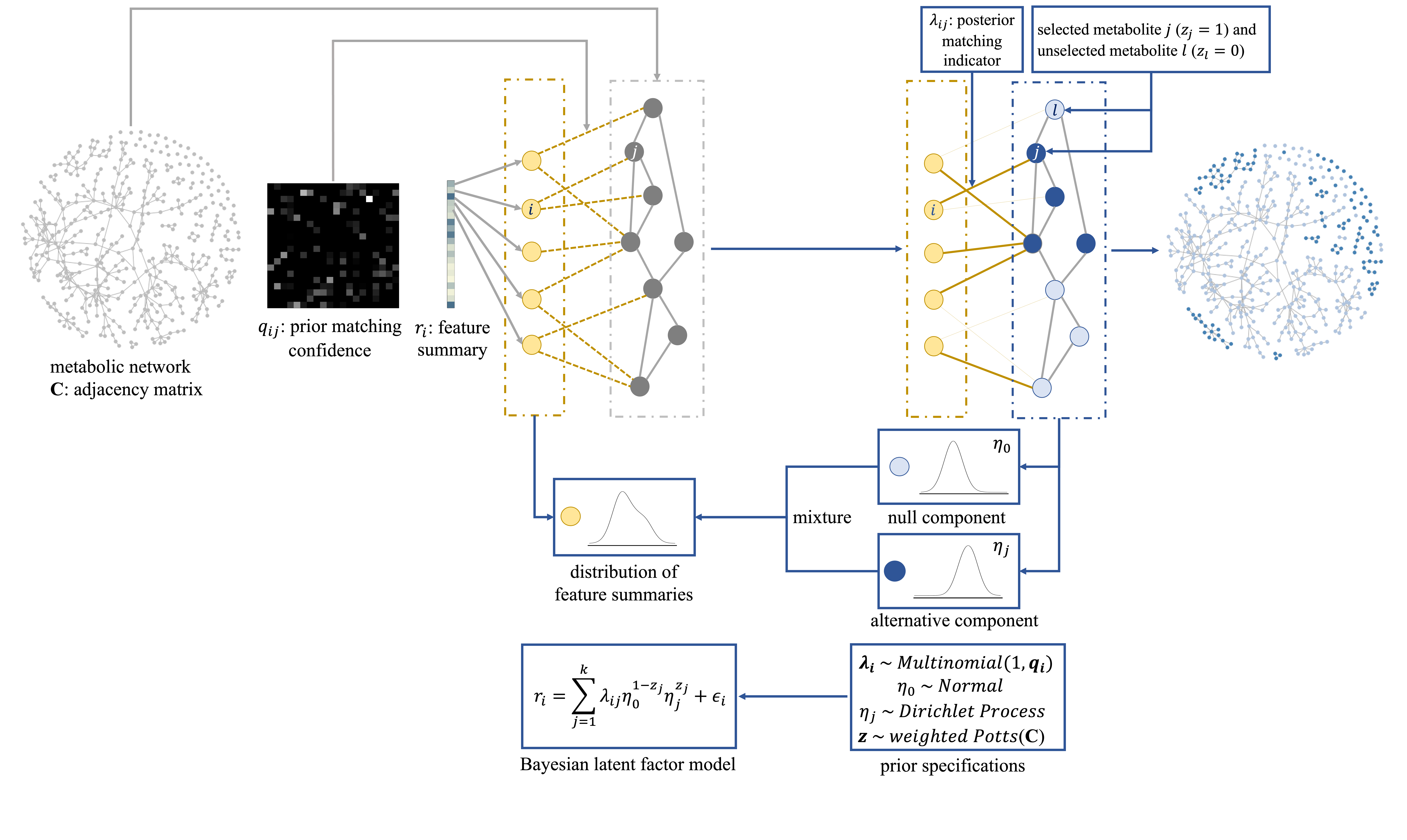}
\caption{The overall setup of BAUM. The observed data for model include the feature-level summary statistics $r_i$ computed from the observed metabolic feature and the clinical outcome (optional), the potential feature-metabolite matches and their confidence measures $q_{ij}$, and the known metabolic network structure. The output of the model include the false discovery rate (FDR) for each metabolite, and the strength of each feature-metabolite matching. We use a Bayesian latent factor model to characterize the observed feature summary statistics and link them to the unobserved metabolite behavior. We assign a Multinomial prior with prior probabilities $\boldsymbol{q}_{i}$ to matching indicators $\boldsymbol\lambda_i$, a normal prior to the null component score $\eta_0$, a Dirichlet Process prior to the alternative component score $\eta_j$ and a weighted Potts prior to metabolite latent class indicators $\boldsymbol z$. Section \ref{sec:methods} provides details of model setup. Generally, the observed summary statistic of a feature is considered to be a linear combination of the unobserved scores of its linked metabolites. The weights reflect the confidence level of the metabolite-feature annotation, and are to be estimated from the data. The metabolites are segregated into two latent classes -- the clinically relevant class (alternative component), and the clinically irrelevant class (null component). The two classes have different distributions of metabolite scores. Metabolites that are connected on the metabolic network are more likely to belong to the same class. }
\label{fig:model}
\end{sidewaysfigure}

We develop a Bayesian constrained latent factor model to characterize the observed feature-level test statistics and link them to the unobserved metabolite behavior and the clinical outcome variable (Figure \ref{fig:model}). Generally, the observed test statistic of a feature is considered to be a linear combination of the unobserved scores of its matched metabolites. The weights reflect the confidence level of the metabolite-feature annotation, and are to be estimated from the data. The metabolites are segregated into two latent classes -- the clinically relevant class, and the clinically irrelevant class. The two classes have different distributions of metabolite scores. Metabolites that are connected on the KEGG metabolic network are more likely to belong to the same class. 

Let $p$ denote the number of observed metabolic features and $k$ denote the number of unobserved metabolites. For $i=1,\dots,p$, $r_i$ denotes the feature-level summary statistics generated by a statistical test that may or may not involve clinical outcomes. For any $i$ and $j$, denote $q_{ij}\in [0, 1]$ the confidence measure of matching feature $i$ to metabolite $j$, which can be calculated based on the multiple matching status and other characteristics of each feature \cite{pmid27977166}. Let $\boldsymbol q_i = (q_{i1}, \ldots, q_{ik})^\top$ and $\sum_{j=1}^k q_{ij}=1$ for all $i$. Denote $\textbf{C} = (c_{jl})$ the adjacency matrix for the metabolic network, where $c_{jl}$ is 1 if there is an edge between metabolite $j$ and metabolite $l$, and 0 otherwise. The observed data include feature-level summary statistics $r_i$, potential feature-metabolite matches and their prior biological confidence measures $\boldsymbol{q}_i$, and the metabolic network structure $\textbf{C}$. We assume the feature-level summary statistics are obtained prior to using BAUM and the feature matrix (and possibly the clinical outcome) is not part of the observed data. The output of the model includes the false discovery rate (FDR) for each metabolite, and the strength of each feature-metabolite matching. 

\subsection{Model}
\subsubsection{One-hot constrained factor analysis model for matching uncertainty}\label{sec:constrained_FA}
We develop a factor analysis approach with one-hot constraints to model the matching uncertainty between observed metabolite features and unobserved metabolites, 
\begin{align}
    r_i = \sum_{j=1}^k \lambda_{ij}\eta_j^* + \epsilon_i, \quad \epsilon_i\overset{i.i.d.}{\sim}\text{N}(0, \sigma^2) \label{eq:matching_uncertainty}
\end{align}
for $i=1, \ldots, p$, where $\eta_j^*$ is the latent score for metabolite $j$ (see Section \ref{sec:latent_mixture}) and $\lambda_{ij}$ is the binary matching indicator between feature $i$ and metabolite $j$, with $\lambda_{ij} = 1$ denoting a match and $\lambda_{ij} = 0$ otherwise. Consequently, for $i=1,\ldots,p$, the observed likelihood for summary statistics is $(r_i \; | \; \eta_j^*, \sigma^2) \sim \text{N}(\eta_j^*, \sigma^2)$ for $j$ such that feature $i$ and metabolite $j$ are matched. Since only one metabolite can be the true match of a feature, for all $i$, we require $\boldsymbol\lambda_i = (\lambda_{i1}, \ldots, \lambda_{ik})^\top$ to be a one-hot binary vector, that is, $\lambda_{ij}=1$ if and only if $\lambda_{il}=0$ for $l \neq j$. Then, with the matching confidence measure $\boldsymbol q_i$, the prior of $\boldsymbol{\lambda}_i$ is $\text{Multinomial}(1, \boldsymbol{q_i})$, where $\lambda_{ij} = 0$ if $q_{ij} = 0$ while $\lambda_{ij}$ may take either 0 or 1 if $q_{ij} > 0$, for all $i=1,\ldots,p$. 

\subsubsection{Mixture model for latent metabolite scores}\label{sec:latent_mixture}

We model the latent metabolite scores $\eta_j^*$ by a mixture distribution of a null component and a alternative component, 
\begin{align}
    \eta_j^*\sim\pi_0 g_0(\eta_j^*) + \pi_1 g_1(\eta_j^*)\label{eq:mixture}
\end{align}
where $\pi_0$ and $\pi_1$ are the proportions of metabolites that are clinically irrelevant and clinically relevant, respectively; functions $g_0$ and $g_1$ represent the densities for the two components. We model the null distribution by a centered Gaussian, i.e., $g_0\sim\text{N}(0, \gamma_0)$, and the alternative component by a Dirichlet process mixture (DPM), i.e., $g_1\sim\mathcal{DP}(P_0, \tau)$, where $\gamma_0$ is the null component variance, $P_0$ is the base measure of a Dirichlet Process defined on $\mathcal{R}\times[0, \infty)$ and $\tau$ is the precision parameter. Equivalently, we can represent this mixture model by
\begin{align}
    \eta_j^* = \eta_{0}^{1-z_j}\eta_{j}^{z_j}, \quad \eta_0 \; | \; \gamma_0 \; \sim \; \text{N}(0, \gamma_0), \quad \eta_j \; \sim \; \mathcal{DP}(P_0, \tau) \label{eq:equvalent_mixture}
\end{align}
where $\eta_j$ is the latent score for clinically relevant metabolite $j$, $\eta_0$ is the latent score for all clinically irrelevant metabolites, and $z_j$ is the binary latent class label for metabolite $j$ (see Section \ref{sec:weigted_potts}). The latent metabolite score $\eta_j^* = \eta_j$ if metabolite $j$ is clinically relevant ($z_j = 1$, the alternative component) and $\eta_j^* = \eta_0$ if metabolite $j$ is clinically irrelevant ($z_j = 0$, the null component). The proportion of clinically relevant metabolites $\pi_0 = \Pr(z_j = 0)$ and the proportion of clinically relevant metabolites $\pi_1=\Pr(z_j = 1)=1-\pi_0$.
%We would like to note that the distribution of latent metabolite scores characterises the distribution of feature summary statistics because (\ref{eq:matching_uncertainty}) implies the feature $i$ is a noisy copy of $\eta_j^*$ (i.e., $r_i=\eta_j^*+\epsilon_i$) if feature $i$ matches metabolite $j$. Since usually $r_i$'s can be obtained by transforming test $p$-values generated into normally distributed statistics where significant features have larger $r_i$'s, 

\subsubsection{Weighted Potts prior for latent metabolite class labels}\label{sec:weigted_potts}

To incorporate the topological structure of the metabolic network, we assign a weighted Potts prior \cite{jin2022feature} to latent metabolite class labels $\boldsymbol{z}=(z_1,\ldots,z_k)^\top$, with the following probability mass function up to a scaling factor,
\begin{align*}
    \pi(\boldsymbol{z}\;|\;\boldsymbol\pi, \boldsymbol\rho, \boldsymbol w, \textbf{C}) \propto \exp\Bigg\{\sum_{j=1}^k\Big[\Tilde{w}_j\log\pi_{z_j}+\rho_{z_j}\sum_{l\ne j}^k w_l c_{lj} \text{I}_{\{z_l=z_j\}}\Big]\Bigg\}
\end{align*}
where $\text{I}_A$ is the indicator function of event $A$, $\boldsymbol\pi = (\pi_0, \pi_1)^\top$, $\boldsymbol\rho = (\rho_0, \rho_1)^\top$ with $\rho_0, \rho_1 \ge 0$, $\boldsymbol w = (w_1, \ldots, w_k)^\top$ with $w_j \ge 0$, and $\textbf{C} = (c_{jl})$ is the metabolic network adjacency matrix. The parameter $\boldsymbol\pi$ is the prior knowledge on the proportions of clinically irrelevant and relevant metabolites. The parameter $\boldsymbol\rho$ controls the global strength of the neighbourhood similarity of $\boldsymbol z$, while weights $\boldsymbol w$ controls the local similarity level. The neighborhood weight $\Tilde{w}_j=\sum_{l=1}^k c_{lj}w_l / \sum_{l=1}^k c_{lj}$ represents the average neighborhood weight for metabolite $j$.

\subsection{Posterior inferences}\label{sec:posterior_inferences}
\subsubsection{Equivalent model representation for latent metabolite scores}\label{sec:equivenlent_representation}
For efficient posterior computation, we construct an equivalent model representation of the alternative component in (\ref{eq:equvalent_mixture}), i.e., $\eta_j \; \sim \; \mathcal{DP}(P_0, \tau)$, by following \cite{neal2000markov} that DPM models can be obtained by taking the limit as the number of clusters goes to infinity. We employ the stick-breaking prior to approximate the Dirichlet Process \cite{ishwaran2001gibbs}. Let $G$ denote the number of clusters, and $\textbf{K}=(K_1,\ldots,K_k)^\top$ denote the cluster labels for metabolites. Let Categorical$(\boldsymbol{p})$ denote the categorical distribution with probabilities $\boldsymbol{p}=(p_1,\ldots,p_G)^\top$ and $\sum_g^G p_g = 1$, i.e., if $K_j\sim\text{Categorical}(\boldsymbol{p})$ then $\Pr(K_j=g)=p_g$, for $g = 1,\dots,G$. Denote $\boldsymbol{m}=(m_1,\ldots,m_G)^\top$ and $\boldsymbol{\gamma}=(\gamma_1,\ldots,\gamma_G)^\top$ as the cluster means and variances, respectively. Then, as $G\rightarrow\infty$, (\ref{eq:equvalent_mixture}) is equivalent to the following, for $j=1,\dots,k$, 
\begin{align*}
    \eta_j \; | \; \boldsymbol m, \boldsymbol\gamma, \textbf{K}\sim\text{N}\left(\sum_{g=1}^G \text{I}_{\{K_j = g\}}m_{g}, \sum_{g=1}^G \text{I}_{\{K_j = g\}}\gamma_{g}\right), \quad
    K_j \; | \; \boldsymbol p\sim\text{Categorical}(\boldsymbol p), %\label{eq:equivalent_DPM}
\end{align*}
where $\boldsymbol{p}\sim\pi(\boldsymbol{s}, \boldsymbol{t})$ is the stick-breaking prior parameterized by $\boldsymbol{s}$ and $\boldsymbol{t}$. 

\subsubsection{Hyperpriors and hyperparameters}

Denote $\text{Gamma}(a, b)$ the Gamma distribution and $\text{IG}(a, b)$ the inverse Gamma distribution, both with shape $a$ and rate $b$. We assign conjugate inverse Gamma priors on the noise variance $\sigma^2$, the null component variance $\gamma_0$ and the alternative component cluster variances $\gamma_g$ for $g=1,\ldots,G$, i.e., $\sigma^2\sim\text{IG}(a_1, b_1)$, $\gamma_0\sim\text{IG}(a_2, b_2)$ and ($\gamma_g\;|\;\beta_g) \sim \text{IG}(a_3, \beta_g)$ for all $g$, where $a_1$, $a_2$, $a_3$, $b_1$ and $b_2$ are hyperparameters, and $\beta_g$ has a conjugate Gamma prior with hyperparameters $a_4$ and $b_4$, i.e., $\beta_g \sim \text{Gamma}(a_4, b_4)$. For the alternative cluster means $m_g$ for all $g$, we assume a normal prior $(m_g \; | \; \sigma^2_g) \sim \text{N}(\mu_g, \sigma^2_g)$ with $\sigma^2_g \sim \text{IG}(a_5, b_5)$ where $\mu_{K_j}$, $a_5$ and $b_5$ are hyperparameters. 

The number of clusters $G$ in the alternative component, the cluster means $\mu_g$ and the proportion of clinically relevant metabolites $\pi_1$ need to be ideally prespecified based on the unknown distribution of metabolite scores. However, in practice, we can use the distribution of feature summary statistics as a close surrogate to the distribution of latent metabolite scores to determine these hyperparameters. This is because $r_i$ is a noisy copy of $\eta_j^*$ (i.e., $r_i=\eta_j^*+\epsilon_i$) when feature $i$ matches metabolite $j$. We suggest $\mu_g$ for $g=1,\ldots,G$ being evenly spaced and cover the range of summary statistics, and then $G$ is determined according to the interval length between two neighbouring $\mu_g$'s and the range of $\mu_g$'s. We set $\pi_1$ to be the proportion of significant features, with $\pi_0 = 1-\pi_1$. While these estimates may not perfectly reflect the unknown distribution of metabolite scores, they provide reasonable prior knowledge. Accurate estimates can be obtained through posterior inferences. For the weighted Potts prior, we set $\rho_0 = \rho_1 = 0.1$ and $w_j=1$ for all $j$. In addition, we advise using a tight prior for the null component, i.e., $a_2 \gg b_2$, to separate it from the alternative component. Because a substantial constitution of BAUM is latent, we recommend assigning informative tight priors to variances, with specific settings described in each application. 

We provide extensive sensitivity analysis demonstrating that BAUM remains very stable under changes in these tight priors in real-world applications. BAUM selects highly consistent pathways across different hyperparameter settings (see Supplementary Materials S3 for details).

\subsubsection{Posterior sampling, parameters of interest, and FDR control}
We develop a blocked Gibbs sampler for posterior inferences. We rely on an equivalent model representation described in Section \ref{sec:equivenlent_representation} for DPM and utilize a blocked Gibbs sampling algorithm for efficient posterior sampling of $\eta_j$'s. We obtain posterior samples of the latent metabolite class labels $\boldsymbol{z}$ by the Swendsen-Wang graph partition algorithm \cite{swendsen1987nonuniversal, jin2022feature} with the weighted Potts prior. Full conditionals for other parameters can be derived in a standard manner. We provide the full conditionals of each parameter and a group updating scheme of $\boldsymbol{z}$ based on the Swendsen-Wang algorithm in Supplementary Materials S1.

The main parameters of interest are $\boldsymbol{z}$ and $\boldsymbol\lambda_i$ for all $i$. We estimate the posterior inclusion probability of metabolite $j$ by the posterior mean of $z_j$. We then control FDR at level $\alpha$ \cite{morris2008bayesian}. Denote the posterior inclusion probabilities as $u_1,\ldots, u_k$. We first sort $\{u_j\}^k_{j=1}$ in descending order to obtain $\{u_{(l)}\}^k_{l=1}$. Then, let $\phi_\alpha = u_{\xi}$ with $\xi = \max\{l^*: (l^*)^{-1}\sum_{l=1}^{l^*} (1-u_{(l)}) \le \alpha \}$. We determine metabolite $j$ as significant if $u_j > \phi_{\xi}$.

We estimate the posterior confidence measure of matching feature $i$ and metabolite $j$ by the posterior mean of $\lambda_{ij}$, denoted by $\hat{\lambda}_{ij}$, and determine feature $i$ matches to metabolite $j$ if $j=\arg\max_l\{\hat{\lambda}_{il}\}$. For all our analyses, we employ a burn-in of 1000 steps followed by 4000 steps for inferences, and we set FDR $\alpha=0.2$. We check convergence by trace plots and auto-regressive correlation plots. 

\subsubsection{Post-processing -- a heuristic approach for quick estimation of metabolite abundance}
We estimate the subject-specific metabolic values are by a convex combination of all features, where the weights are based on the estimate of matching uncertainty (e.g., the posterior mean of $\lambda_{ij}$). Specifically, denote the posterior mean of $\lambda_{ij}$ by $\lambda_{ij}^*$, and $\boldsymbol\lambda_j^* = (\lambda_{1j}, \lambda_{2j}, ..., \lambda_{nj})^\top$. Then, we can estimate the metabolite $j$ abundance of subject $s$ by $\hat{m}_{sj}=c\boldsymbol\lambda_j^{*T}\boldsymbol{x}_s$ where $\boldsymbol{x}_s\in\mathbb{R}^n$ is the value of the $n$ features for subject $s$, and $c$ is a scaling factor such that the weights of feature values sum up to 1.  

\section{Simulations}

We perform extensive and realistic simulations to evaluate the performance of BAUM, varying a number of network specifications, including feature count $p$, metabolite count $k$, the alternative component, unmatched metabolite percentage, potential feature-metabolite matchings, and the metabolite network structures. We consider four simulation scenarios, two based on generative networks (\textbf{GN1} and \textbf{GN2}) and two based on real-world feature-metabolite networks (\textbf{RN1} and \textbf{RN2}). Table \ref{tab:configuration} summarizes the key differences between these scenarios. In \textbf{GN1} and \textbf{GN2}, we set $p=k=1000$ and generate metabolite networks using the Barabasi-Albert model \cite{albert2002statistical}. The alternative component's metabolite scores follow $\mN(10, 1)$, and potential feature-metabolite matches are random. \textbf{GN1} ensures every metabolite has at least one potential matching to features, while \textbf{GN2} introduces 50\% unmatched metabolites to mimic real-world conditions. For \textbf{RN1} and \textbf{RN2}, we utilize the network from the COVID-19 metabolomics data \cite{coviddata, sindelar2021longitudinal} used in Section \ref{sec:results_covid-19} with $p=1153$ features and $k=1093$ metabolites. The metabolite network used in \textbf{RN1} and \textbf{RN2} contains 13\% metabolites without potential matches. In \textbf{RN1}, the alternative component is $\mN(10, 1)$ while in \textbf{RN2} is $\chi^2(10)$. In all scenarios, metabolite labels are determined based on their vertex degrees, with higher-degree metabolites more likely to belong to the alternative component. The alternative component's metabolite scores may vary, but the null component always has a score of zero. Finally, feature-level summary statistics are generated according to (\ref{eq:matching_uncertainty}).

\begin{table}
\centering
\caption{Differences in the four simulation scenarios. In both \textbf{GN1} and \textbf{GN2}, there are $p=1000$ features and $k=1000$ metabolites, and the alternative components (AC) are $\mN(10, 1)$. The metabolite networks are generated by simulating scale-free networks. The percentage of unmatched metabolites (\% UM) is 50\% in \textbf{GN2} while every metabolite has potential matchings to features in \textbf{GN1}. In both \textbf{RN1} and \textbf{RN2}, potential feature-metabolite matchings and the metabolite networks are obtained from the ST001849 COVID-19 metabolomics data, where we use $p=1153$ features and $k=1093$ human metabolites. The metabolite network has 13\% metabolites with no potential matchings to any features. In \textbf{RN1}, the alternative component is $\mN(10, 1)$ while in \textbf{RN2} is $\chi^2(10)$. Abbreviations in table: AC -- alternative component, \% UM -- percentage of unmatched metabolites. }
\begin{tabular}{r c c c c c}
    \hline\hline
    \textbf{Settings}   & $\boldsymbol p$   & $\boldsymbol k$   & \textbf{Network}  & \textbf{AC}     & \textbf{\% UM} \\
    \hline
    \textbf{GN1}        & 1000              & 1000              & Scale-free        & $\mN(10, 1)$    & 0\%   \\
    \textbf{GN2}        & 1000              & 1000              & Scale-free        & $\mN(10, 1)$    & 50\%  \\
    \textbf{RN1}        & 1153              & 1093              & COIVD-19          & $\mN(10, 1)$    & 13\%  \\
    \textbf{RN2}        & 1153              & 1093              & COIVD-19          & $\chi^2(10)$    & 13\%  \\
\hline\hline
\end{tabular}
\label{tab:configuration}
\end{table}

For \textbf{GN1}, \textbf{GN2} and \textbf{RN1}, we utilize a single cluster ($G=1$) to simplify the alternative component as a Gaussian distribution. The mean of the alternative component ($m_1$) is degenerate at 10. In contrast, for \textbf{RN2}, we employ 21 clusters ($G=21$) for the alternative component. These clusters have prior means ($\mu_g$) taking integers from 5 to 25 based on feature summary statistics. For \textbf{RN2}, we specify $a_5=1\text{e}4$ and $b_5=1$, while $a_5$ and $b_5$ remain unspecified in the other scenarios since $m_1$ is degenerate. Common parameters across all scenarios include $a_1=2\text{e}4$, $a_2=a_3=a_4=b_1=1\text{e}4$ and $b_2=b_4=1$. Using a histogram of feature-level statistics, we determine $\pi_1 = 0.15$ for \textbf{GN1} and \textbf{GN2}, $\pi_1=0.2$ for \textbf{RN1} and $\pi_1=0.4$ for \textbf{RN2}. We repeat each scenarios for 100 times. 

\begin{table}[t!]
\centering
\small
\spacingset{1}
\caption{Simulation results for two generative network scenarios (\textbf{GN1} and \textbf{GN2}) and two real-network scenarios (\textbf{RN1} and \textbf{RN2}) on (a) metabolite inferences and (b) matching estimations. Summary statistics Mean (s.d.) are based on 100 replicates.}
\begin{subtable}{\textwidth}
    \centering
    \caption{Metabolite inferences in different simulation scenarios and for different methods. We compare BAUM with LocFDR and Post-LocFDR. }
    \begin{tabular}{r r c c c c}
    \hline\hline
    \textbf{Settings}             & \textbf{Methods}    & \textbf{ACC}  & \textbf{AUC}  & \textbf{FPR}  & \textbf{TPR}  \\
    \hline
    \multirow{3}{*}{\textbf{GN1}} & \textbf{BAUM}        & 95.1\% (0.8\%)& 93.3\% (1.4\%)& 4.3\% (0.7\%) & 91.0\% (2.7\%)\\
                                  & \textbf{LocFDR}     & 91.0\% (1.0\%)& 92.9\% (2.2\%)& 9.6\% (1.2\%) & 95.3\% (4.5\%)\\
                                  & \textbf{Post-LocFDR}& 95.6\% (0.8\%)& 91.4\% (1.4\%)& 3.0\% (0.8\%) & 85.9\% (2.7\%)\\
    \hline
    \multirow{3}{*}{\textbf{GN2}} & \textbf{BAUM}        & 99.1\% (0.5\%)& 97.4\% (1.4\%)& 0.3\% (0.3\%) & 95.0\% (2.8\%)\\
                                  & \textbf{LocFDR}     & 88.6\% (2.2\%)& 93.5\% (1.3\%)& 13.0\% (2.6\%)& 99.9\% (0.4\%)\\
                                  & \textbf{Post-LocFDR}& 95.3\% (2.5\%)& 95.8\% (1.7\%)& 4.8\% (2.9\%) & 96.5\% (2.5\%)\\
    \hline
    \multirow{3}{*}{\textbf{RN1}} & \textbf{BAUM}        & 96.7\% (0.6\%)& 91.7\% (1.3\%)& 0.8\% (0.4\%) & 84.1\% (2.6\%)\\
                                  & \textbf{LocFDR}     & 36.7\% (2.0\%)& 62.0\% (1.2\%)& 76.0\% (2.4\%)& 100\% (0\%)\\
                                  & \textbf{Post-LocFDR}& 83.7\% (3.0\%)& 83.7\% (2.8\%)& 16.3\% (3.6\%)& 83.8\% (2.4\%)\\
    \hline
    \multirow{3}{*}{\textbf{RN2}} & \textbf{BAUM}        & 94.4\% (1.1\%)& 94.2\% (1.2\%)& 4.9\% (1.4\%) & 93.3\% (2.0\%)\\
                                  & \textbf{LocFDR}     & 79.9\% (1.9\%)& 83.4\% (1.5\%)& 32.0\% (3.0\%)& 98.9\% (0.7\%)\\
                                  & \textbf{Post-LocFDR}& 85.1\% (1.9\%)& 87.7\% (1.5\%)& 23.8\% (3.2\%)& 99.2\% (0.6\%)\\
    \hline\hline
\end{tabular}
\label{tab:simulation_meta}
\end{subtable}
\vfill
\vspace{0.2cm}
\begin{subtable}{\textwidth}
    \centering
    \caption{Matching estimation results for BAUM in different simulation scenarios.}
    \begin{tabular}{r c c c c}
    \hline\hline
    \textbf{Settings}   & \textbf{ACC}  & \textbf{AUC}  & \textbf{FPR}  & \textbf{TPR}  \\
    \hline
    \textbf{GN1}        & 96.4\% (0.9\%)& 91.2\% (2.0\%)& 85.7\% (4.0\%)& 2.2\% (0.8\%) \\
    \textbf{GN2}        & 97.8\% (0.6\%)& 97.2\% (1.1\%)& 96.5\% (2.1\%)& 2.1\% (0.7\%) \\
    \textbf{RN1}        & 94.2\% (0.9\%)& 87.0\% (2.1\%)& 74.8\% (4.2\%)& 0.7\% (0.2\%) \\
    \textbf{RN2}        & 88.6\% (1.4\%)& 87.3\% (1.5\%)& 79.3\% (2.6\%)& 4.7\% (1.2\%) \\
    \hline\hline
    \end{tabular}
    \label{tab:simulation_match}
\end{subtable}
\label{tab:simulation}
\end{table}

We evaluate the metabolite selection performance on matched metabolites since unmatched metabolites do not directly contribute to the data likelihood, making their latent class inferences highly challenging. For matching estimation, instead of checking if BAUM identifies true feature-metabolite matches, we assess its capability to match the features correctly to the null component or to the alternative component, because it is intrinsically difficult to distinguish between potential matches when a feature has potentially matches to multiple metabolites with similar latent scores. 

Table \ref{tab:simulation_meta} presents a comparison between BAUM and local FDR control (LocFDR) for metabolite selections. We also include the performance of LocFDR after incorporating BAUM's matching uncertainty estimation (Post-LocFDR). There are methods of Bayesian local FDR control \cite{jin2022feature} in literature but such methods are not designed for the matching uncertainty estimation and can encounter numerical issues. For LocFDR, we assume all metabolites potentially matched to a feature have the same matching probability. We compute metabolite-specific statistics as weighted averages of feature statistics for features potentially matched to the metabolite. For Post-LocFDR, we compute metabolite-specific statistics using the posterior probability of matching as weights. Metabolite selections are based on the metabolite-specific statistics while controlling the local FDR at 0.2. Table \ref{tab:simulation_match} shows the feature-metabolite matching performance only for BAUM, as it is the first method to quantify the feature-metabolite matching uncertainty. 

As shown in Table \ref{tab:simulation_meta}, across all simulations, BAUM consistently achieves superior metabolite selection results. In challenging scenarios with real-world networks, BAUM maintains a good and stable performance. In some scenarios, LocFDR exhibits higher true positive rate (TPR) at the expense of much higher false positive rate (FPR). Both LocFDR and Post-LocFDR display lower accuracy (ACC), lower area under the curve (AUC) and higher FPR in metabolite selections than BAUM, which shows the importance of jointly modeling matching uncertainty and metabolite behaviour in enhancing metabolite selections. Notably, Post-LocFDR shows substantial improvement over LocFDR in accuracy, AUC, and FPR, indicating the informativeness and utility of our matching estimation on metabolite-level inferences. Table \ref{tab:simulation_match} shows the feature-metabolite matching estimated by BAUM accurately distinguishes the null component and the alternative component for the metabolites, maintaining good accuracy and AUC, even in scenarios with complex real-world networks. In all scenarios, matching FPR is well-controlled.

\section{Results}

We analyzed a COVID-19 metabolomics dataset and a mouse brain development metabo-lomics dataset using BAUM. BAUM finds pathways that are consistent with existing knowledge, as well as novel pathways that are biologically plausible. We provide hyperparameters used in each application in Supplementary Materials S2. Furthermore, sensitivity analysis showed that BAUM was very robust and selected highly consistent pathways acorss different hyperparameter settings, even with a very small sample size of 16 mice per group in the analysis of the mouse brain data (please see Supplementary Materials S3 for details). 

\subsection{COVID-19 metabolomics data}\label{sec:results_covid-19}
We analyzed the COVID-19 metabolomics data \cite{coviddata, sindelar2021longitudinal} using BAUM. The dataset (ST001849) was downloaded from the NIH Metabolomics Workbench, which was derived from an untargeted metabolomics study of subjects who were positive of SARS-CoV-2. The purpose was to find indicators of disease severity using baseline metabolomics at admission. After removing subjects with unknown ICU admission status, a total of 269 subjects who were SARS-CoV-2 positive were studied, among which 133 were admitted to ICU. Using day-0 metabolomics data, our analysis tried to find metabolic signatures that can separate those who were admitted to ICU from the less severe cases. 

The dataset contained subject-level observations of 5471 metabolic features (3819 positive-ion features and 1652 negative-ion features). After matching the features to metabolites on the KEGG network \cite{KEGG}, we kept 1153 features that had at least one match to the human metabolic network and 1093 human metabolites. The kept metabolites are either directly linked to features or part of path between metabolites that are linked to features. Among the 1153 features we studied, 582 (50.5\%) were matched to only one metabolite, hence no matching uncertainty was involved. However, 151 (13.1\%) features had at least 5 possible matches to metabolites. From the metabolite perspective, among the 1093 metabolites, 142 (13.0\%) metabolites had no match to any features, 334 (30.6\%) were matched to only one feature, and 217 (20.0\%) had at least 5 feature matches.

We performed marginal distance correlation t-tests \cite{szekely2014partial} to detect non-linear associations between features and the binary ICU status, and used the resulting t-statistics as the feature-specific summary statistics. %We dropped all 69 subjects with missing ICU admission status.

\begin{figure}[ht!]
\centering 
\small
\spacingset{1}
\includegraphics[width=16cm]{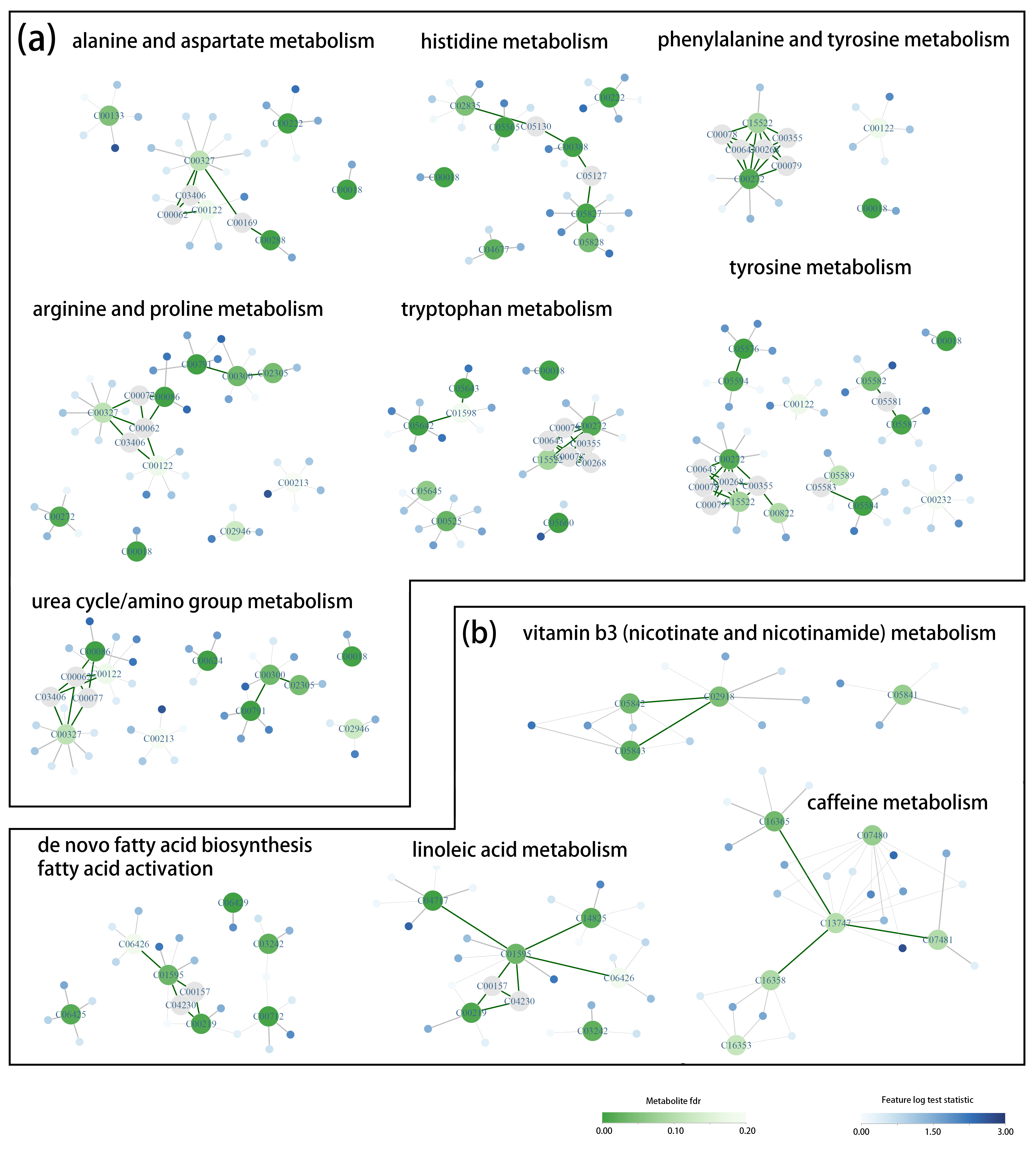}
\caption{Selected subnetworks and their corresponding pathways. Green nodes: selected metabolites at FDR $\leq 0.$2; blue nodes: features matched to the metabolites; gray nodes: metabolites that are not selected, but connect selected metabolites. (a) Subnetworks associated with amino acid and nucleotide metabolism. (b) Other subnetworks. }
\label{fig:covid}
\end{figure}

We controlled FDR based on the posterior probability of whether the metabolite is clinically relevant to the outcome by adopting the procedure described in \cite{morris2008bayesian}. Among the 1093 metabolites, BAUM selected 189 clinically relevant metabolites to the outcome by controlling FDR at level 0.2. 

Figure \ref{fig:covid} shows the selected sub-networks from the human metabolic network. For each sub-network, the most significant pathway(s) were found using pathways in the metapone package \cite{RN100tian} and the hypergeometric test for over-representation \cite{beissbarth2004gostat}. The majority of the sub-networks were part of the central metabolism of amino acids and nucleotides (Figure \ref{fig:covid}a). It has been found that COVID infection alters amino acid metabolism \cite{pmid35064792}, and the level of changes are linked to disease severity \cite{pmid34502454}. Besides general amino acid metabolism changes, some specific amino acids were clearly linked to the physiology of COVID infection. We found tyrosine and tryptophan metabolism to be associated with ICU admission. Other studies have found tyrosine metabolic pathway was prominently affected by COVID infection in oral secretion samples, after correcting for stress response of the immune system \cite{pmid36690957}. The imbalance of the urea cycle can cause severe inflammatory damage. It has been reported that ornithine concentration is higher in critically ill COVID patients. At the same time, arginine concentration is significantly lower, and arginine-ornithine conversion dominates the urea cycle in COVID patients \cite{jia2022metabolomic}. The degradation of arginine leads to the accumulation of its downstream metabolites and exacerbates the inflammatory response. The increase of aspartate and its downstream product asparagine provides a favorable environment for the translation of viral mRNA \cite{chatterjee2006arginine}. 

Another interesting pathway is the caffeine metabolism pathway. Though a commonplace diet component, caffeine has been shown to be an anti-inflammatory chemical, as well as an immuno-modulator, with a specific effect on airway smooth muscle. It is believed to achieve this function by acting as a phosphodiesterase inhibitor and adenosine receptor antagonist \cite{pmid34067243}. Among the selected metabolites was caffeine (C07481) itself. In addition, 1-Methylxanthine (C16358), AFMU (C16365), and paraxanthine (C13747) were also selected. Paraxanthine is known for its attenuation effect on the formation of cholestatic liver fibrosis \cite{pmid22032678}. 

The linoleate metabolism pathway was selected. It has been found that increased serum linoleic acid was associated with more severe symptoms of COVID \cite{pmid35502320}. Interestingly, a recent structural study suggested that linoleic acid can bind to the spike glycoprotein of the coronavirus, potentially exerting anti-viral effect \cite{pmid36762857}. In the dataset under study, linoleic acid level is lower on average in the group with more severe symptom. Among the other selected metabolites in the linoleate metabolism pathway was arachidonic acid (C00219), which is known to be an endogenous antiviral metabolite, the lack of which can make the the person less resistant to the coronavirus \cite{pmid32583353}. The average abundance of arachidonic acid was also lower in the more severe group in the current dataset. 

The vitamin B3 (nicotinate and nicotinamide) metabolism pathway was also selected by our method. It has been documented that the vitamine B3 pathway, together with tryptophan metabolism pathwasy mentioned above, is altered in severe SARS-COV2 patients \cite{pmid33712622}. A nutritional intervention with nicotinamide was beneficial when combined with other therapy for coronavirus. A clinical trial has been conducted by NIH (NCT04751604) to study the impact of vitamin B3 on the disease course of COVID-19. Interestingly, the related compound nicotine was also considered potentially beneficial in SARS-COV2 resistance, potentially through the mechanism of nicotinic acetylcholine receptor (nAChR). 

\subsection{Mouse brain data - development and healthy aging in different brain regions}\label{sec:results_mice}

We analyzed the mouse brain atlas data (ST001637), which was downloaded from the NIH Metabolomics Workbench \cite{micedata, ding2021metabolome}. The dataset consisted of 480 mice split into 60 groups. The groups were characterized by two genders (male and female), three age points (3, 16 and 59 weeks) and ten brain regions. Each group had eight mice subjects.

The dataset had observations of 17032 metabolic features (10085 positive- and 6947 negative- ion features). %We extracted a mice-specific metabolic network of size 3095 from the full KEGG metabolite pathway database. 
After matching the features to known mouse metabolites, we kept 819 features that had at least one match to the mouse metabolites. We screened out 950 mice metabolites that neither matched to any features nor was along the path between two metabolites which matched to features. Among the 819 kept features, 548 (67.0\%) were matched to only one features, while 46 (5.6\%) has at least 5 feature matches. Among the 2145 mice metabolites, 1207 (56.3\%) had no match to any features, 598 (27.9\%) were matched to one feature and 7 (0.3\%) had at least 5 matched features.

Given the small sample size in each age/gender/brain region combination, we focused on linear relations between metabolites and development/aging. We employed the analysis of variance (ANOVA) to detect metabolic difference among mice of different age groups, in Hippocampus, Olfactory Bulb and Thalamus, respectively. We made two age group comparisons -- week 3 v.s. week 16 (development), and week 16 v.s. week 59 (healthy aging). For each of the brain region, we performed an ANOVA for each feature controlling for gender. Then we transformed the $F$-statistic values by normal quantile transformation, and took the transformed statistics as the feature-level summary.

\begin{figure}[t]
\centering 
\small
\spacingset{1}
\includegraphics[width=16cm]{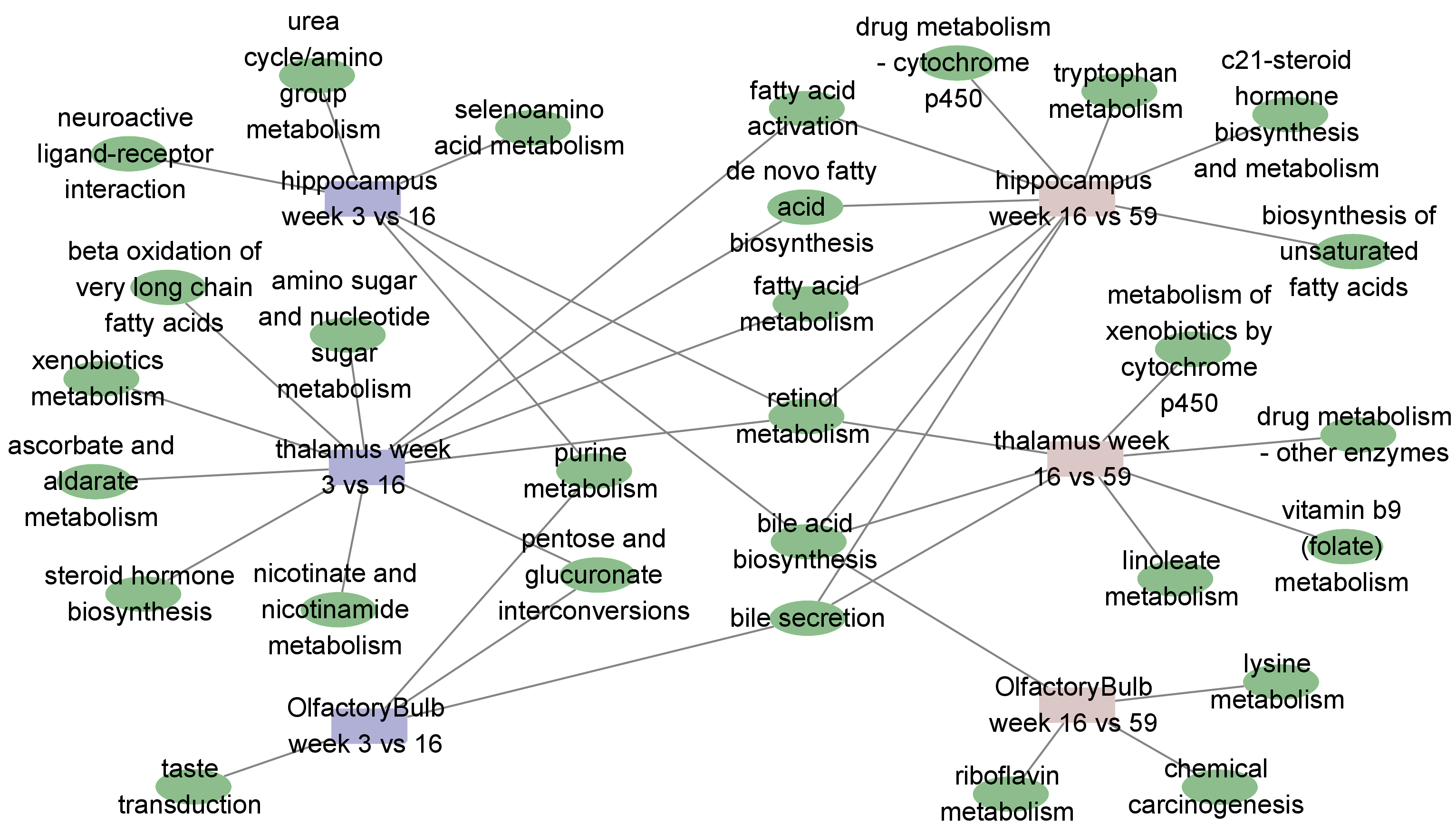}
\caption{Mouse brain data results: significant pathways that are associated with different stages and brain regions.}
\label{fig:mouse1}
\end{figure}

After selecting significant metabolites at $\text{FDR} \leq 0.2$ in each comparison group, again the most significant pathway(s) were found using pathways in the metapone package \cite{RN100tian} and the hypergeometric test for over-representation. Given we are making separate comparisons for different brain regions and different age group contrasts, and given the small sample size in each comparison (16 vs 16) yielding small number of significant metabolites, we conducted the pathway analysis using all significant metabolites for each comparison. We selected pathways with $\geq 3$ significant metabolites, as well as with $p$-values $\leq 0.05$. Figure \ref{fig:mouse1} shows the selected pathways connected with the 6 comparisons.

\begin{figure}[ht!]
\centering 
\small
\spacingset{1}
\includegraphics[width=16cm]{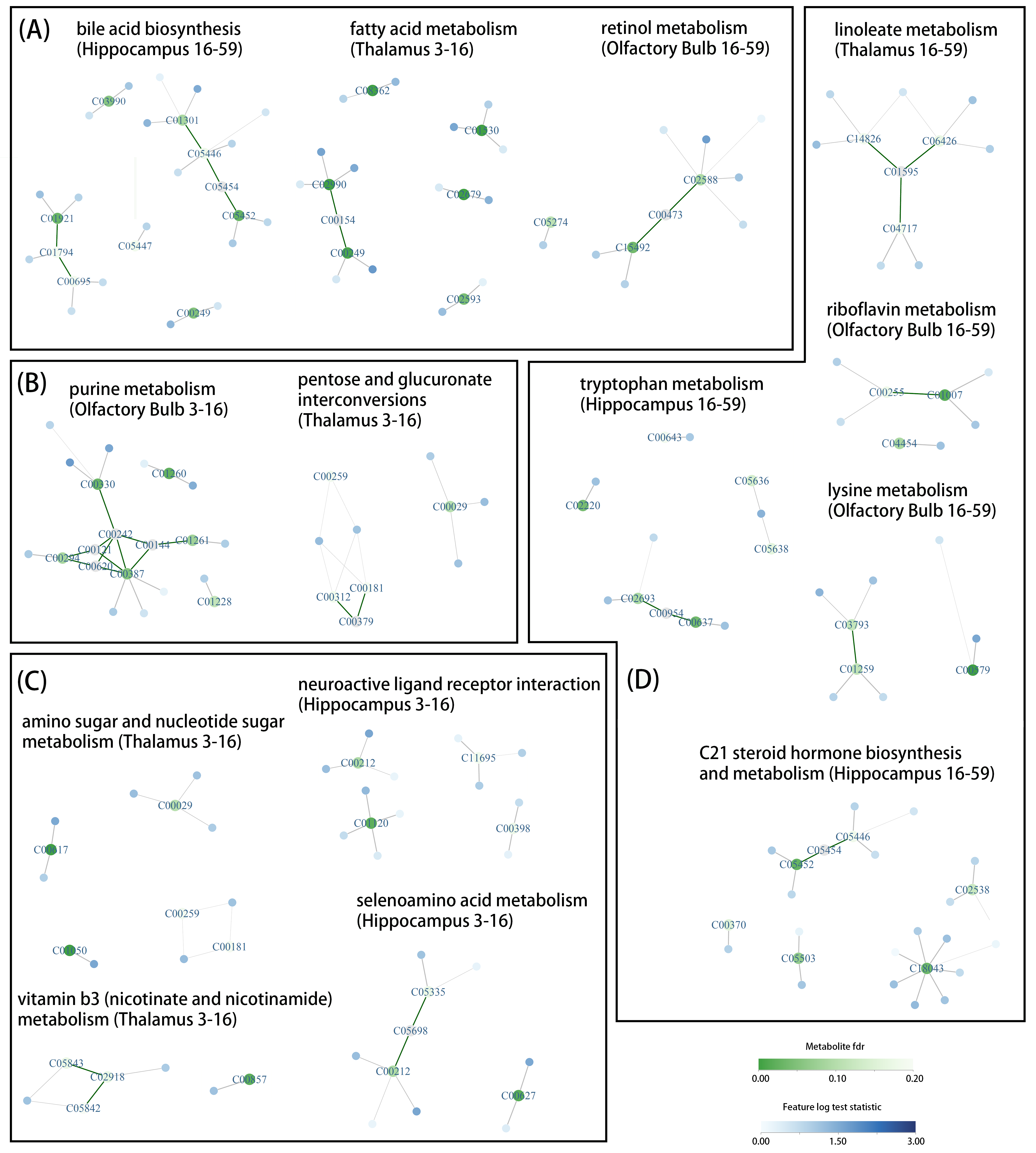}
\caption{Example significant pathways from the mouse brain data. Green nodes: selected metabolites at FDR $\leq 0.2$; blue nodes: features matched to the metabolites; gray nodes: metabolites that are not selected, but connect selected metabolites. (A) Pathways connected to both development and healthy aging; (B) Pathways connected to development of multiple brain regions; (C) Example pathways connected to development of certain brain regions; (D) Example pathways connected to healthy aging of certain brain regions.}
\label{fig:mouse2}
\end{figure}

Figure \ref{fig:mouse2} shows details of some example pathways. Some pathways are significant in multiple brain regions or age group comparisons. For such pathways we selected a single brain region and age group comparison that involves the largest number of significant metabolites as example. Figures for all selected pathways in all age group comparisons are in the Supplementary Materials S4.

The most notable pathway in Figure \ref{fig:mouse1} and Figure \ref{fig:mouse2}(A) is the retinol metabolism pathway, which is associated with both hippocampus and thalamus, in both development and aging stages. Retinoids are well known for their important role in nervous system development, as well as the development of many other bodily structures \cite{pmid16688755}. Retinyl palmitate (C02588) is a common form of retinol derivative in the brain. In the current data, its level is substantially higher in hippocampus and thalamus in adult mice than in baby mice. Supplementation of retinyl palmitate appears to have disruptive effects in developing and adult rat brain \cite{pmid32726121}. 

%C00473 Retinol; C02588 Retinyl palmitate; C15492 all-trans-13,14-Dihydroretinol (part of all-trans retinyl ester) "RA signaling regulates development of many organs and tissues, including the body axis, spinal cord, forelimbs, heart, eye and reproductive tract" (PMID: 31273085) "Retinoids (vitamin A) are crucial for most forms of life. In chordates, they have important roles in the developing nervous system and notochord and many other embryonic structures, as well as in maintenance of epithelial surfaces, immune competence, and reproduction. " Retinol is one of the retinoids. (PMID: 16688755)\\

Another group of pathways that is well-known in brain function is the fatty acid metabo-lism pathways (Figure \ref{fig:mouse1} and Figure \ref{fig:mouse2}(A)). Fatty acid derivatives influence many brain functions \cite{pmid28065757}. The pathways show strong relations with thalamus at the developmental stage, and hippocampus in the healthy aging stage. However we also noticed that some of the metabolites are also linked to the brain regions in other stages. The significant metabolites include L-Palmitoylcarnitine (C02990), Hexadecanoic acid (C00249), Palmitoleic acid (C08362), Stearic acid (C01530), Lauric acid (C02679), Tetradecanoyl-CoA (C02593), Decanoyl-CoA (C05274) etc. Among the metabolites found in this study, steric acid (C01530) was identified as a potential marker for AD and aging in a human study \cite{pmid33718349}. Lauric acid showed beneficial effects in neuronal maturation and neuroprotection against oxidative stress in cellular and animal models \cite{pmid32420479}.

Two bile acid pathways are widely connected with multiple brain regions in development and aging (Figure \ref{fig:mouse1} and Figure \ref{fig:mouse2}(A)). Bile acids are cholesterol-derived steroid acids that serve as signaling molecules mostly for nutrient availability \cite{pmid35338368}. The key enzyme in the pathway, CYP7A1, also plays an important role in the clearance of brain cholesterol. Bile acids CA, DCA, and CDCA are able to influence neurotransmission \cite{pmid27468758}, and play important roles in gut-liver-brain axis and normal brain functions. Disruption of gut microbiome can cause neurological disorder through bile acid signaling \cite{pmid36400238}. The selected metabolites in this pathway include 3$\alpha$,7$\alpha$,12$\alpha$,26-Tetrahydroxy-5$\beta$-cholestane (C05446), 3$\alpha$,7$\alpha$,12$\alpha$-Trihydroxy-5$\beta$-cholestane (C05454), 3$\alpha$,7$\alpha$,12$\alpha$-Trihydroxy-5$\beta$-cholestan-26-al (C01301), 3$\alpha$,7$\alpha$-Dihydroxy-5$\beta$-cholestane (C05452), Glycocholic acid (C01921), Palmitic acid (C00249), Lithocholic acid (C03990). CUrrently mechanistic studies linking the bile acids and their derivatives to brain development and aging is still scarce. Studies have found that palmitic acid has the potential to trigger neuroinflammation in the brain \cite{pmid34131233}, and the serum level of lithocholic acid, as well as two other bile acids deoxycholic acid and glycoursodeoxycholic acid, are higher in Alzheimer's patients \cite{pmid36994114}. 

Figure \ref{fig:mouse2}(B) shows two pathways connected to the development stage. As an example, purines and their derivatives are centrally involved in energy homeostasis and DNA synthesis. In addition, purinergic signalling plays critical roles in the nervous system \cite{pmid26056033}. Disruptions to purine metabolism can cause a multitude of neurological disorders \cite{pmid30441833}. The selected metabolites include Guanosine (C00387), Deoxyguanosine (C00330), Inosine (C00294), P1,P4-Bis(5'-guanosyl) tetraphosphate (C01261), P1,P4-Bis(5'-adenosyl) tetraphosphate (Ap4A; C01260), and Guanosine 3',5'-bis(diphosphate) (C01228). Among them, guanosine is a known neuromodulator \cite{pmid27699087}. Studies have shown that inosine and Ap4A have neuroprotective effects \cite{pmid27761681}. 

Figure \ref{fig:mouse2}(C) shows pathways connected to a single brain region either in development. Besides the self-explanatory pathway "neuroactive ligan receptor interaction", in which selected metatolites include adenosine (C00212), sphinganine 1-phosphate (C01120), anandamide (C11695) and tryptamine (C00398), another interesting example is the selenoamino acid metabo-lism pathway. The impact of selenoproteins and selenoamino acids has been mostly studied from the perspective of neurodegenerative diseases and food supplements \cite{pmid33762907}. It has been found that selenomethionine promotes hippocampus neurogenesis in AD \cite{pmid28109879}. In the current study, the level of selenomethionine (C05335) decreases over the development of the hippocampus.  

Figure \ref{fig:mouse2}(D) show pathways connected to a single brain region in healthy aging. As an example, the riboflavin metabolism pathway is associated with healthy aging of olfactory bulb. Ribovlavin is a water-soluble B vitamin that protects against oxidative stress. Its metabolism is of critical importance in brain health, the dysfunction of which can contribute to neurodegenerative disease \cite{pmid33886098}. 

Another well-known pathway is C21 steroid hormone biosynthesis and metabolism. Derivatives of two of the three members of the estrogen family, estrone and estradiol were found to be significant. They include Estrone-3-sulfate (E1S, C02538) and Estradiol 3-glucuronide (E2-3G, C05503). E1S is an inactive form of E1, which can be taken up by cells to synthesize E2 \cite{pmid23476785}. It is a major form of estrone involved in brain-blood efflux transport of estrogens. E2 is known to be associated with aging and menopausal transition, as well as some neurodegenerative disorders, potentially through the cholinergic and dopaminergic systems \cite{pmid31364065}. Another selected metabolite, Cholesterol sulfate (C18043) is a critical component of the cell membrane, which serves to stabilize membrane structure. It has a neuroprotective effect by reducing oxidative stress \cite{pmid31425677}. Overall, our method was able to select informative metabolites and pathways, even under very small sample size of 16 mice per group. 

%Not talked about in C21: 
%C05452     3alpha,7alpha-Dihydroxy-5beta-cholestane; 5beta-Cholestane-3alpha,7alpha-diol
%C05446   3alpha,7alpha,12alpha,26-Tetrahydroxy-5beta-cholestane;
%C00370          Sterol
%-----------------------------------------------------------------------------------------
%pentose
%C00029     UDP-glucose;
%C00312     L-Xylulose	
%C00181     D-Xylose
%C00259     L-Arabinose

%amino sugar
%C00617      UDP-D-galacturonate;
%C01050      UDP-N-acetylmuramate;
%C00029      UDP-glucose;

%vitamin b3
%articles/PMC4772032
%C05843       N1-Methyl-4-pyridone-5-carboxamide;
%C05842       N1-Methyl-2-pyridone-5-carboxamide
%C02918       1-Methylnicotinamide
%C00857       Deamino-NAD+;

%linoleate

%C14826    12(13)-EpOME;
%C06426  gamma-Linolenic acid;Gamolenic acid
%C04717  (9Z,11E)-(13S)-13-Hydroperoxyoctadeca-9,11-dienoic acid;13(S)-HPODE;
%C01595  Linoleate;

%tryptophan
%C02220  2-Aminomuconate;
%C02693  (Indol-3-yl)acetamide; Indole-3-acetamide
%C00637  Indole-3-acetaldehyde; Indoleacetaldehyde
%C00643  5-Hydroxy-L-tryptophan; Oxitriptan
%C05636  3-Hydroxykynurenamine
%C05638  5-Hydroxykynurenamine

%lysine
%C03793      N6,N6,N6-Trimethyl-L-lysine
%C01259      (3S)-3-Hydroxy-N6,N6,N6-trimethyl-L-lysine
%C00579      Dihydrolipoamide;

\section{Discussion}
In this study, we designed a framework Bayesian Analysis for Untargeted Metabolomics data (BAUM), which can select important metabolites that tend to be functionally consistent, make inference on the matching between metabolites and observed features, and incorporate non-linear associations between features and the clinical outcome. BAUM utilizes the existing knowledge graph of the relations between metabolites in the form of a metabolic network. A drawback is that BAUM ignores data features that do not match to known metabolites in the network. On the other hand, as most studies focus on core metabolic pathways, BAUM is powerful in that it can make statistical inference on the metabolites' association with the outcome and feature-metabolite matching simultaneously, partially resolving the issue of multiple matching. The Bayesian framework makes BAUM very robust, which can make inferences based on small sample sizes, solving a challenge in many metabolomics studies. We used BAUM to analyze two real datasets and obtained biologically meaningful results.  

%\section*{Author Contributions}
%J.K., T.Y. and G.M. conceived the study. G.M. and J.K. developed the method. G.M. conducted numerical experiments and analyzed the data. T.Y. took primary responsibility to interpret the results in real data analysis. All authors drafted and reviewed the entire paper. 

\section*{Key points}
\begin{itemize}
    \item We develop a innovative approach for Bayesian Analysis of Untargeted Metabolomics Data (BAUM) to integrate previously separate tasks into a single framework, including matching uncertainty inference, metabolite selection, and functional analysis. 
    \item BAUM can identify subnetworks within the entire metabolic network based on feature-level summary statistics, enhancing biological interpretation.
    \item Under the Bayesian framework, BAUM is robust and stable, and can make inferences based on small sample sizes.
    \item Simulations show BAUM can make accurate inferences on the feature-metabolite matchings and metabolite significance. 
    \item BAUM finds pathways that conform to existing knowledge as well as novel pathways that are biologically plausible on two real-world dataset. 
\end{itemize}

\section*{Data availability}
The COVID-19 metabolomics dataset (ST001849) was downloaded from the NIH Metabo-lomics Workbench at \url{https://www.metabolomicsworkbench.org/data/DRCCMetadata.php?Mode=Study&StudyID=ST001849}. The mouse brain atlas data (ST001637) was downloaded from the NIH Metabolomics Workbench at \url{https://www.metabolomicsworkbench.org/data/DRCCMetadata.php?Mode=Study&StudyID=ST001637}. The preprocessed data of these two datasets are available at \url{https://github.com/guoxuan-ma/BAUM}. 

\section*{Code availability}
We provide an R package ``BAUM" for analyzing untargeted metabolomics data by our method. The R package is available at \url{https://github.com/guoxuan-ma/BAUM}. 

\section*{Fundings}
This work was partially supported by the National Key R\&D Program of China grant 2022ZD0116004, NIH grants (R01DA048993, R01MH105561, R01GM124061), and NSF grant IIS2123777.

\clearpage

\bibliographystyle{mystyle}
\bibliography{ref}

\end{document}